\documentclass{article} 

\usepackage{epsfig}
\usepackage{latexsym}

\begin{document}

\vspace*{10 mm}

\begin{center}
{\Large\bf Axions, Quantum Mechanical Pumping,\\[5mm]
and Primeval Magnetic Fields}\\[22mm]
{\large J\"urg Fr\"ohlich$\;^{1,2\ast}$ }, \,
{\large Bill Pedrini$\;^{1\dag}$ } \\[7mm]
$^1\;$ Institut f\"ur Theoretische Physik \\
ETH H\"onggerberg\\ CH\,--\,8093\, Z\"urich\\[5mm]
$^2\;$ I.H.E.S. \\
35, Rte de Chartres,\\ F-91440 Bures-sur-Yvette\\[7mm]
$^{\ast}\;$\textsf{juerg@itp.phys.ethz.ch}\\[3mm]
$^{\dag}\;$\textsf{pedrini@itp.phys.ethz.ch}x

\end{center}

\vspace{25 mm}

\begin{abstract}
We discuss the ordinary quantum Hall effect and a higher-dimensional
cousin. We consider the dimensional reduction of these effects to $1+1$ and
$3+1$ 
space-time dimensions, respectively. After dimensional reduction, an axion
field appears, which plays the r\^{o}le of a space-time dependent difference
of chemical potentials of chiral modes. As applications, we sketch a theory of
quantum pumps and a mechanism for the generation of primeval magnetic fields
in the early universe.
\end{abstract}

\newpage

\section{Introduction}
\begin{figure}[ht]
\vskip.2in
\begin{center}
\input{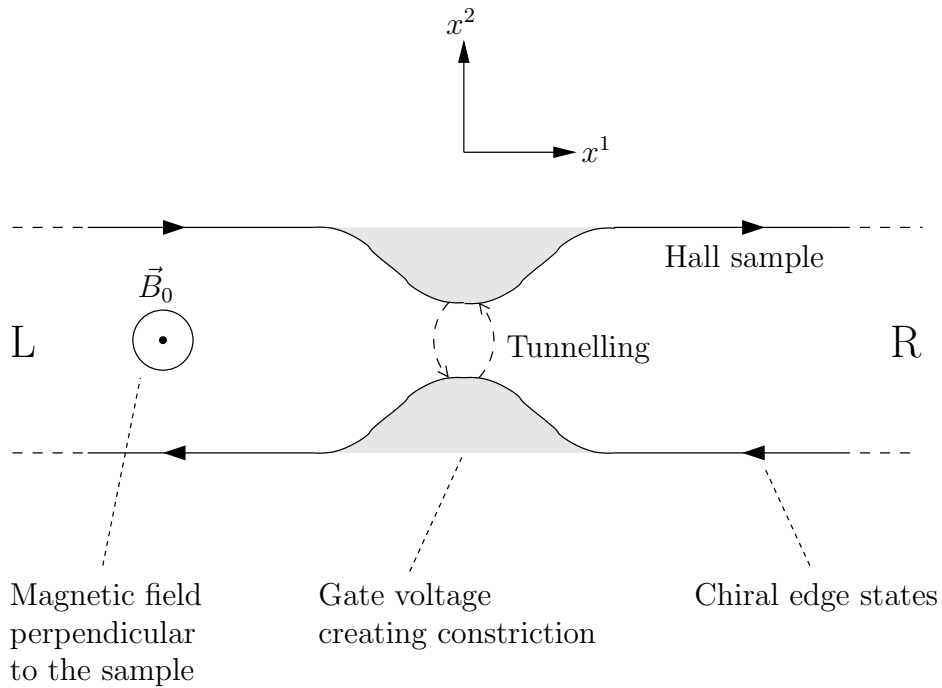}
\caption{\label{fig:constriction}A constriction in a quantum Hall sample.}
\end{center}
\end{figure}

In these notes, we clarify the r\^{o}le played by certain pseudo-scalar fields
related to ``axions'' in some transport- or pumping processes in semiconductor
devices and in the early universe. These processes are similar to ones
observed in quantum Hall systems. We therefore start by recalling some key
features of the theory of the quantum Hall effect. We then consider transport
processes in very long, narrow rectangular Hall samples with constrictions, as
shown in Figure \ref{fig:constriction}.

For samples of this kind, filled with an incompressible Hall fluid, the
component, $A_2$, of the electromagnetic vector potential, $A$, parallel to the
short axis, $2$, of the sample can be interpreted as a pseudo-scalar field
analogous to the \textit{axion} known from elemantary particle physics
\cite{axions}. In the region where the sample has a constriction,
\textit{tunnelling processes} between the chiral edge modes on the upper and
lower edge of the sample may occur. It is interesting to consider the effect
of turning on a time-dependent voltage drop in the $2$-direction. Not
surprisingly, we find that when such a voltage drop, $V(t)$, with
\begin{equation}
  \int_{-\infty}^{+\infty}V(t)dt=:\delta\varphi
\nonumber
\end{equation}
is turned on, an electric charge $\delta Q$ proportional to $\delta\varphi$ is
transported through the constriction from the left, $L$, to the right,
$R$. This system thus realizes a simple ``\textit{quantum pump}''. Due to the
tunnelling processes between edge states of opposite chirality, the
state of the pump exhibits a periodicity in $\delta\varphi$ proportional to
the inverse electric charge of the charge carriers in the sample. Thus, such a
pump can be used, in principle, to explore properties of the quasi particles
in incompressible quantum Hall fluids, such as their electric charges, \cite{saclay}.

Our model can also be used to describe quantum wires carrying a Luttinger
liquid. The r\^{o}le of the constriction is then played by impurities mixing
left and right movers.

We then proceed to studying a five-dimensional analogue of the quantum Hall
effect. If four-dimensional physics is described by dimensional reduction from
a five-dimensional slab to two parallel boundary ``3-branes'', the axion can
be interpreted as the component of the five-dimensional electromagnetic vector
potential transversal to the branes. Tunnelling of chiral fermions from one to
the other brane, due e.g. to a mass term, generates a periodic axion
potential. It is then argued that the dynamics of the axion may trigger the
growth of large-scale primeval magnetic fields in the early universe. In other
words, axion dynamics - which is coupled to the dynamics of the curvature
tensor of space-time - can be viewed as a realization of a quantum-field
theoretical ``pump'' driving the growth of large-scale primeval magnetic
fields, \cite{tkachev}\cite{tuwi}\cite{jubi}; see also \cite{josha}. Whether
this mechanism plays a r\^{o}le in explaining the observed large-scale
magnetic fields in the universe is, however, still uncertain; see \cite{giovagraru}.

\section{Brief Recap of the Quantum Hall Effect}
  \label{sec:recap}

We consider a uniform $2$-dimensional electron gas of density $n$ forming at
the interface between a semiconductor and an insulator when a gate voltage is
applied in the direction perpendicular to the interface. We imagine that a
homogeneous magnetic field, $\vec{B}_0$, perpendicular to the interface is
turned on. Let $\nu:=(nhc)/(e|\vec{B}_0|)$ denote the ``filling factor''. From
the experiments of von Klitzing et al. \cite{vonkli} and Tsui et
al. \cite{tsui} we have learnt that, for certain values of $\nu$, the
$2$-dimensional electron gas forms an incompressible fluid, in the sense that
the longitudinal resistence, $R_L$, of the system vanishes.
We consider the response of such a system to turning on a small external
electromagnetic field $(\underline{E},B)$, where $\underline{E}$ denotes the
in-plane component of the electric field, and $B_{\mathrm{tot}}=B_0+B$ is the
component of the total magnetic field perpendicular to the plane of the fluid.
By $\underline{j}(x)$ we denote the current density in the plane of the
$2$-dimensional electron gas, and by $j^0(x)=\rho(x)+en$ the deviation of the
electric charge density from the uniform background charge density, $en$;
(here $x=(\underline{x},t)$, where $\underline{x}$ is a point in the sample
and $t$ is time).

By combining Hall's law (for $R_L=0$), i.e.,
\begin{equation} \label{eq:hall}
  j^k(x)=G_H\varepsilon^{kl}E_l(x)
  \quad,
\end{equation}
where $G_H$ is the bulk Hall conductance, with the continuity equation for
$j_0$ and $\underline{j}$ and Faraday's induction law, one easily finds that 
\begin{equation} \label{eq:charge}
  j^0(x)=G_HB(x)
  \quad,
\end{equation}
see \cite{froke}. Denoting by $F=(F_{\mu\nu})$ the electromagnetic field tensor
over the $(2+1)$-dimensional space-time, $\Lambda$, of the sample and by
$J\equiv(J_{\mu\nu}=\varepsilon_{\mu\nu\lambda}j^{\lambda})$ the $2$-form
dual to the charge-current density $(j^0,\underline{j})$, eqs. (\ref{eq:hall}) and
(\ref{eq:charge}) can be summarized in
\begin{equation} \label{eq:chargecurrent}
  J=G_HF
  \quad,
\end{equation}
the field equation of ``Chern-Simons electrodynamics'' \cite{dejate}. Defining
the dimensionless Hall conductivity, $\sigma_H$, by
\begin{equation} \label{eq:sigmahall}
  \sigma_H=\frac{h}{e^2}G_H\chi_{\Lambda}(x)
\end{equation}
and using units such that $e^2/h=1$, the field equations of Chern-Simons
electrodynamics are
\begin{equation} \label{eq:csfieq}
  J(x)=\sigma_H(x)F(x)
  \quad.
\end{equation}
Taking the exterior derivative of eq. (\ref{eq:csfieq}), we find that
\begin{equation} \label{eq:divcsfieq}
  dJ=d\sigma_H\wedge F
  \quad,
\end{equation}
because $dF=d(dA)=0$. The gradient $d\sigma_H$ is transversal to the boundary,
$\partial\Lambda$, of the sample's space-time. Eq. (\ref{eq:divcsfieq})
 tries to tell us
that electric charge is \textit{not} conserved in an incompressible Hall
fluid, because $dJ$, the dual of
$\partial_{\mu}j^{\mu}=\partial_tj^0+\mathrm{div}\underline{j}$, does
\textit{not} vanish. The origin of this false impression is that, so far, we
have neglected the \textit{diamagnetic edge current}, $J_{\mathrm{edge}}$, in
our equations. This current is localized on $\partial\Lambda$ and is dual to a
vector field $i=(i_{\mu})$ with support on $\partial\Lambda$ and parallel to
$\partial\Lambda$. The edge current $J_{\mathrm{edge}}$ saves electric charge
conservation:
\begin{equation} \label{eq:charcons}
  d(J+J_{\mathrm{edge}})=0
  \quad.
\end{equation}
Eqs. (\ref{eq:charcons}) and (\ref{eq:divcsfieq}) then yield
\begin{equation} \label{eq:divbc}
  \partial_{\mu}i^{\mu}=\sigma_H^{\mathrm{edge}}E^{\parallel}
  \quad,
\end{equation}
where $E^{\parallel}$ is the component of the electric field parallel to the
boundary of the sample, and the ``edge'' conductivity,
$\sigma_H^{\mathrm{edge}}$, is equal to $-\sigma_H$, the ``bulk''
conductivity, as follows from (\ref{eq:divcsfieq}). Eq. (\ref{eq:divbc})
describes the $(1+1)$-dimensional chiral anomaly \cite{jack}. Apparently, the
edge current, $i$, is an anomalous (chiral) electric current localized on the
boundary of the sample; (the chirality of $i$ depends on the direction of
$\vec{B}_0$ and the sign of the electric charge of the fundamental charge
carriers).

Equations (\ref{eq:divbc}) and (\ref{eq:csfieq}) can be derived from an action
principle. If $S^{\mathrm{eff}}_{\Lambda}(A)$ denotes the \textit{effective
  action}, i.e., the generating functional of the current Green functions, of
an incompressible Hall fluid confined to a three-dimensional space-time region
$\Lambda$, in the presence of an external electromagnetic field
$(\underline{E},B)$ with vector potential $A$, then
\begin{equation} \label{eq:effac}
  S^{\mathrm{eff}}_{\Lambda}(A)\approx
  \frac{\sigma_H}{2}\int_{\Lambda}A\wedge dA+
  \Gamma_{\partial\Lambda}(a)
  \quad,
\end{equation}
where $a=\left.A^{\parallel}\right|_{\partial\Lambda}$ is the restriction of
$A$ to the boundary, $\partial\Lambda$, of $\Lambda$, and ``$\approx$'' means
that only the \textit{leading contributions} (in the sense of dimensional
analysis) to the effective action are displayed on the R.S. The first (bulk)
term on the R.S. of (\ref{eq:effac}) is the \textit{Chern-Simons action}, the
second (edge) term turns out to be the \textit{anomalous chiral action}
\cite{jack} in two space-time dimensions. Its gauge variation fixes the value
of $\sigma_H^{\mathrm{edge}}$ by
\begin{equation}
  \left.\frac{d}{d\varepsilon}\right|_{\varepsilon=0}
  \Gamma_{\partial\Lambda}(a+\varepsilon d\chi)=
  \chi\;\sigma_H^{\mathrm{edge}}E^{\parallel}
  \quad.
\end{equation}
Electromagnetic gauge invariance is
a fundamental property of non-relativistic many-body theory. Thus,
$S^{\mathrm{eff}}_{\Lambda}(A)$ must be gauge invariant, i.e.,
\begin{equation} \label{eq:effacc}
  S^{\mathrm{eff}}_{\Lambda}(A)=S^{\mathrm{eff}}_{\Lambda}(A+d\chi)
  \quad,
\end{equation}
for an arbitrary function $\chi$ on $\Lambda$. Individually, the Chern-Simons
action,\\ $\frac{\sigma_H}{2}\int_{\Lambda}A\wedge dA$, and the boundary action
$\Gamma_{\partial\Lambda}(a)$ are \textit{not} invariant under a gauge
transformation, $\chi$, \textit{not} vanishing on the boundary
$\partial\Lambda$; but the R.S. of (\ref{eq:effac}) \textit{is} gauge
invariant precisely if $\sigma_H=-\sigma_H^{\mathrm{edge}}$.

Since $S^{\mathrm{eff}}_{\Lambda}(A)$ is the generating functional of the
current Green functions, we have that
\begin{equation} \label{eq:currgreen}
  j^{\mu}(x)=\frac{\delta S^{\mathrm{eff}}_{\Lambda}(A)}{\delta A_{\mu}(x)}
  \quad,\quad
  i^{\mu}(x)=
  \frac{\delta \Gamma_{\partial\Lambda}(a)}{\delta a_{\mu}(x)}
  \quad.
\end{equation}
These expressions, togheter with eq. (\ref{eq:effac}) for
$S^{\mathrm{eff}}_{\Lambda}(A)$, reproduce the basic equations
(\ref{eq:csfieq}) and (\ref{eq:divbc}).

The boundary action $\Gamma_{\partial\Lambda}(a)$ is known to be the
generating functional of the chiral Kac-Moody current operators of current
algebra with gauge group $U(1)$. It is then a natural idea
\cite{jualle}\cite{jjcb} that the boundary degrees of freedom of an
incompressible Hall fluid are described by a \textit{chiral conformal field
  theory}. Under the natural assumptions that
\begin{itemize}
\item[(i)] sectors of physical states of this theory are labelled by their
  electric charge and, possibly, finitely many further quantum numbers
  (e.g. spin) with finitely many possible values; and
\item[(ii)] excitations of this theory with even/odd electric charge (in units
  where $e=1$) obey Bose/Fermi statistics,
\end{itemize}
one shows that $\sigma_H$ is necessarily a \textit{rational number}, and one
obtains a table of values of $\sigma_H$ that compares well with those of the
dimensionless Hall conductivity of experimentally established incompressible
Hall fluids, \cite{jualle}\cite{jjcb}. Moreover, one can systematically work
out the spectrum of fractionally charged quasi-particles propagating along the
edge of the sample. The smallest fractional electric charge turns out to be
given by $q=er/d_H$, where $r$ is a positive integer - and, for many fluids,
$r=1$ - and $d_H$ is the integer denominator of $\sigma_H$, (writing
$\sigma_H=n_H/d_H$, with $n_H$ and $d_H$ relatively prime integers); see
\cite{jualle}\cite{jjcb}.

\section{Hall Samples with Constriction \\ and Quantum Wires}
  \label{sec:constriction}

In this section we consider a very long, narrow rectangular Hall sample, as
shown in Figure \ref{fig:constriction}.
The axis parallel to the long side of the sample is taken to be the $1$-axis,
the one parallel to the short side is the $2$-axis, and we set
$\underline{x}=(x^1,x^2)\equiv(x,y)$. We define a field $\varphi$ by
\begin{equation} \label{eq:phi}
  \varphi(x,t)=\int_{\gamma_{\mathtt{lu}}}A_2(x,y,t)dy
  \quad,
\end{equation}
where $\mathtt{l}=(x,y_{\mathtt{l}})$ is a point on the lower edge of the
sample, $\mathtt{u}=(x,y_{\mathtt{u}})$ is a point on the upper edge, and
$\gamma_{\mathtt{lu}}$ is the straight line from $\mathtt{l}$ to
$\mathtt{u}$. We assume that the $1$-component,
\begin{equation}
  E\equiv E_1=\partial_0A_1-\partial_1A_0
  \quad,
\end{equation}
of the in-plane electric field, $\underline{E}$, is independent of $y$. It is
convenient to choose a gauge such that
$A\equiv(A_0,A_1)$ is independent of $y$. Then
the effective action in equation (\ref{eq:effac}) becomes
\begin{equation} \label{eq:effac1}
  S^{\mathrm{eff}}(\varphi,A)\approx
  \sigma_H\int dt \int_{I}dx \;\varphi E
  \quad,
\end{equation}
where $I$ is the interval on the $x$-axis which the Hall sample is confined
to. The terms corresponding to the upper and the lower edge in the boundary
action $\Gamma_{\partial\Lambda}$ on the R.S. of (\ref{eq:effac}) cancel each
other, because $A_0,A_1$ are independent of $y$, \textit{up to a manifestly
  gauge-invariant term} proportional to $\int dt\int dx (A^T)^2$. The action
$S^{\mathrm{eff}}(\varphi,A)$ describes the coupling of an ``\textit{axion
  field}'' $\varphi(x,t)$ to the electric field $E(x,t)$ of
$(1+1)$-dimensional QED. For the current, $I$, through the sample and the
charge density, $P$, in an external axion field configuration $\varphi$, we
find the expressions
\begin{equation} \label{eq:I}
  I(x,t)=\frac{\delta S^{\mathrm{eff}}(\varphi,A)}{\delta A_1(x,t)}
        =-\sigma_H\dot{\varphi}(x,t)
  \quad,
\end{equation}
\begin{equation} \label{eq:P}
  P(x,t)=\frac{\delta S^{\mathrm{eff}}(\varphi,A)}{\delta A_0(x,t)}
        =\sigma_H\varphi'(x,t)
  \quad,
\end{equation}
provided $E=0$, so that there are no contributions from the boundary
action. (Here $I(x,t)=\int j^1(x,y,t)dy$, $P(x,t)=\int j^0(x,y,t)dy$. We
observe that (\ref{eq:I}) and (\ref{eq:P}) imply the continuity equation
$\dot{P}+I'=0$.)

The action $S^{\mathrm{eff}}(\varphi,A)$ in eq. (\ref{eq:effac1}) yields an
accurate description of charge transport in a long, narrow sample filled with
an incompressible Hall fluid with Hall conductivity $\sigma_H$ if the electric
field in the $1$-direction vanishes (so that the term proportional to $\int
dt\int dx (A^T)^2$ does not contribute), as long as tunnelling processes
between the upper and the lower edge can be neglected. However, for a sample
with a constriction, as shown in Figure \ref{fig:constriction}, such tunnelling
processes \textit{do} occur. In a description of the Hall fluid in terms of an
action that displays the edge degrees of freedom explicitly, tunnelling
processes between the two edges are described by terms of the form 
\begin{eqnarray} 
  \int & \left[
    t(x)\sum_{\alpha}
    \bar{\psi}_{\mathrm{left},\alpha}(x,y_{\mathtt{u}},t)
    e^{2\pi i q_{\alpha}\int_{y_{\mathtt{l}}}^{y_{\mathtt{u}}}A_2(x,y,t)dy}
    \psi_{\mathrm{right},\alpha}(x,y_{\mathtt{l}},t) \right] dxdt 
     &+ \nonumber\\
    \quad\quad
    &+ \mathrm{h.c.}\;(\mathrm{left}\leftrightarrow\mathrm{right})
  \quad, &\label{eq:tunnel}
\end{eqnarray}
where $\alpha$ labels the different species of charged quasi-particles
described by left chiral fields,
$\psi_{\mathrm{left},\alpha},\bar{\psi}_{\mathrm{left},\alpha}$, on the upper
edge and by right chiral fields
$\psi_{\mathrm{right},\alpha},\bar{\psi}_{\mathrm{right},\alpha}$, on the lower
edge, and $q_{\alpha}e$ is the electric charge of a quasi-particle of species
$\alpha$. Setting 
\begin{equation} \label{eq:fermifields}
  \stackrel{(-)}{\psi}_{L,\alpha}(x,t)=
  \stackrel{(-)}{\psi}_{\mathrm{left},\alpha}(x,y_{\mathtt{u}},t)
  \quad,\quad
  \stackrel{(-)}{\psi}_{R,\alpha}(x,t)=
  \stackrel{(-)}{\psi}_{\mathrm{right},\alpha}(x,y_{\mathtt{l}},t)
  \quad,
\end{equation}
and recalling eq. (\ref{eq:phi}), the term (\ref{eq:tunnel}) can be written as
\begin{equation} \label{eq:tunnel1}
  \int \left[
    t(x)\sum_{\alpha}
    \bar{\psi}_{L,\alpha}(x,t)
    e^{2\pi i q_{\alpha}\varphi(x,t)}
    \psi_{R,\alpha}(x,t)
    +\mathrm{h.c.}\;(L\leftrightarrow R)
  \right]dxdt
  \quad.
\end{equation}
The function $t(x)$ is a measure for the strength of the amplitude of
tunnelling between the two edges; $|t(x)|$ is ``large'' for $x$ close to the
constriction, and tends to $0$ rapidly, as the distance of $x$ to the
constriction increases.

Besides (\ref{eq:tunnel1}), the action for the edge degrees of freedom
contains terms not mixing the left- and right-moving degrees of freedom. These
terms do \textit{not} depend on $\varphi$. Integrating (or ``tracing'') out
all edge degrees of freedom, we obtain an effective ``boundary action'',
$\tilde{\Gamma}(\varphi,A\equiv(A_0,A_1))$, which now depends on $\varphi$ ! It
is \textit{periodic} in $\varphi$: if $\varphi_0$ is the smallest real number
such that
\begin{equation} \label{eq:period}
  q_{\alpha}\varphi_0=n_{\alpha}
  \quad,\quad
  n_{\alpha}\in\mathbf{Z}
  \quad,
\end{equation}
for all species $\alpha$, then
\begin{equation} \label{eq:periodicity}
  \tilde{\Gamma}(\varphi(\cdot,\cdot)+\varphi_0,A)=
  \tilde{\Gamma}(\varphi(\cdot,\cdot),A)
  \quad.
\end{equation}
This follows immediately from the form of (\ref{eq:tunnel1}) of the tunnelling
terms in the boundary action.

The remarks on the relation between fractional charges and the value of the
Hall conductivity $\sigma_H$ at the very end of Section \ref{sec:recap} lead
to the equation
\begin{equation} \label{eq:sigmaphi}
  \sigma_H\varphi_0=\frac{n_H}{r}
  \quad,
\end{equation}
where $n_H$ is the Hall numerator and $r$ is an integer, (see \cite{jualle};
actually $r=1$, for the Laughlin- and the simple Jain fluids with
$\sigma_H=n/(2pn+1)$, $p,n=1,2,\ldots$).

The total effective action is given by
\begin{equation} \label{eq:effac2}
  S^{\mathrm{eff}}(\varphi,A)\approx
  \sigma_H\int dt \int_{I}dx \;\varphi E
  \;+\;\tilde{\Gamma}(\varphi,A)
  \quad.
\end{equation}
The periodicity property (\ref{eq:periodicity}) of $\tilde{\Gamma}(\varphi,A)$
implies that, if the $1$-component of the electric field vanishes $E=0$, 
the macroscopic state of this system depends
periodically on the external ``axion field'' $\varphi$, with period
$\varphi_0$, and that eqs. (\ref{eq:I}) and (\ref{eq:P}) for the electric
current $I(x,t)$ and the 
charge density $P(x,t)$ continue to hold \textit{in average} when the system
is driven through several cycles. 
Indeed, because of the invariance of $\tilde{\Gamma}(\varphi,A)$ under a
gauge transformation $A'=A+d\chi$, one has
\begin{equation}
  \partial_t\frac{\delta}{\delta A_0(x,t)}\tilde{\Gamma}(\varphi,A)\;+\;
  \partial_x\frac{\delta}{\delta A_1(x,t)}\tilde{\Gamma}(\varphi,A)\;=\;0
  \quad,
\end{equation}
and one can write
\begin{equation}
  \frac{\delta}{\delta A_1(x,t)}\tilde{\Gamma}(\varphi,0)=
  \partial_t U(\varphi,x,t)
  \quad,
\end{equation}
where the function $U(\varphi,x,t)$, which depends on the axion field
configuration $\varphi$ and the spacetime point $(x,t)$, is given by
\begin{equation}
  U(\varphi,x,t)=
  -\int_{-\infty}^x dy 
    \;\frac{\delta}{\delta A_0(y,t)}\tilde{\Gamma}(\varphi,0)
  \quad.
\end{equation}
The function $U$ is periodic in $\varphi$, with period $\varphi_0$,
\begin{equation}
  U(\varphi(\cdot,\cdot)+\varphi_0,x,t)=U(\varphi(\cdot,\cdot),x,t)
  \quad,
\end{equation}
and does not depend on time explicitely,
\begin{equation}
  U(\varphi(\cdot,\cdot+\Delta t),x,t+\Delta t)
  =U(\varphi(\cdot,\cdot),x,t)
  \quad.
\end{equation}
Consider a pump which works with a period $T$, i.e., a pump driven by an axion
field $\varphi(\cdot,\cdot)$ which fulfills
\begin{equation}
  \varphi(\cdot,\cdot+T)=\varphi(\cdot,\cdot)+n\varphi_0
\end{equation}
for some integer $n$. 
One then finds that the charge transport due to 
the second term on the R.H.S. of (\ref{eq:effac2}) vanishes, since
\begin{equation}
  \int_{t}^{t+T}dt\frac{\delta}{\delta A_1(x,t)}\tilde{\Gamma}(\varphi,0)
  \;=\;
  U(\varphi(\cdot,\cdot+T),x,t+T)-U(\varphi(\cdot,\cdot),x,t)\;=\;0
  \quad.
\end{equation}

We now recall the physical meaning of the axion field $\varphi$. By
eq. (\ref{eq:phi}),
\begin{equation} \label{eq:phidot}
  \dot{\varphi}(x,t)=\int_{\gamma_{\mathtt{lu}}}\dot{A}_2(x,y,t)dy=
                     \int_{\gamma_{\mathtt{lu}}}E_2(x,y,t)dy=V(x,t)
  \quad,
\end{equation}
where $V(x,t)$ is the \textit{voltage drop} at $x$ between the lower and the
upper edge of the sample; (we are using that
$E_2=\partial_0A_2-\partial_2A_0=\partial_0A_2$, because $A_0$ is independent
of $y$). Let $\varphi(t)$ be an $x$-independent configuration of the ``axion
field'', with
\begin{equation} \label{eq:deltaphi}
  \delta\varphi:=\int_{-\infty}^{+\infty}\dot{\varphi}(t)dt
                =\int_{-\infty}^{+\infty}V(t)dt
  \quad.
\end{equation}
Then eq. (\ref{eq:I}) tells us that the total amount, $\delta Q$, of electric
charge transported from the left $(L)$ to the right $(R)$ of the sample is
given by
\begin{equation} \label{deltaq}
  \delta Q =\int_{-\infty}^{+\infty}I(x,t)dt=-\sigma_H\delta\varphi
  \quad,
\end{equation}
(and $P(x,t)\equiv0$, by eq. (\ref{eq:P})). Thus, a sample with a
time-dependent voltage drop between the upper and lower edge can be viewed as
a ``quantum pump'' transporting electric charge from the left to the
right. The macroscopic state of this pump is periodic in $\delta\varphi$ with
period $\varphi_0$. Thus, when the pump is operated over $n=1,2,\ldots$
cycles, a total amount, $\delta Q_n$, of electric charge 
\begin{equation} \label{eq:totcharge}
  \delta Q_n=-\sigma_Hn\varphi_o=-\frac{nn_H}{r}
\end{equation}
is transported from the left to the right; (here we have used
eq. (\ref{eq:sigmaphi})). Since $\varphi_0^{-1}$ is the smallest fractional
electric charge of a quasiparticle tunnelling through the constriction, a
measurement of this charge can be obtained from independent measurement of the
charge $\delta Q_n$ transported from the left to the right in $n$ cycles and
of $\sigma_H$. Whether a given voltage pulse
$\delta\varphi=\int_{-\infty}^{+\infty}V(t)dt$ corresponds to an integer
number of cycles of the pump can be inferred from the \textit{fluctuations} of
the charge, $\delta Q$, transported from the left to the right during that
pulse around its mean value $-\sigma_H\delta\varphi$: if, on the left and
right ends the sample is connected to free-electron leads then (independently
of $\delta\varphi$) $\delta Q$ must be an integer (multiple of $e$). If
$\sigma_H\delta\varphi$ is \textit{not} an integer then $\delta Q$ will
exhibit fluctuations around its mean value $-\sigma_H\delta\varphi$. But if
$\delta\varphi$ corresponds to exactly $nr$ cycles, $n=1,2,\ldots$, then
$-\sigma_H\delta\varphi=nn_H$ is an integer, and hence the fluctuations of
$\delta Q$ in this process will essentially \textit{vanish}.

Typical features of the effective action $\tilde{\Gamma}(\varphi)$, with
$E=0$, can be determined by measuring the tunnelling current $I_T$ through the
constriction: when $E=0$
\begin{equation} \label{eq:tunnelcurr}
  I_T(t)=\int dx\lambda(x)
         \frac{\delta\tilde{\Gamma}(\varphi)}{\delta\varphi(x,t)}
  \quad,
\end{equation}
where $\lambda(x)$ is the width of the sample at $x$. A tunnelling current
$I_T$ can be generated e.g. by a modulation of the magnetic field
perpendicular to the plane of the sample. The expression for the tunnelling
current in terms of $\tilde{\Gamma}$ given above shows that, from measurements
of $I_T$ and of the voltage drop $V$ as functions of time, one can infer the
period $\varphi_0$ of $\tilde{\Gamma}$ and, hence, the smallest fractional
electric charge of the quasi-particles. Furthermore, one can argue that the
fluctuations of $I_T$ are proportional to the fractional charge of the
quasi-particles tunnelling through the constriction - an effect used in the
experiments described in \cite{saclay} to measure the
fractional charges of quasi-particles.

If the magnetic field is set to $0$ our considerations can also be used to
describe \textit{quantum wires}. Then $e\dot{\varphi}(x,t)$ has the
interpretation of a (space-time dependent) \textit{difference of chemical
  potentials} between left- and right-moving modes in the
wire. Eq. (\ref{eq:I}) then says that if there isn't any chirality-reversing
scattering in the wire (i.e. $\tilde{\Gamma}(\varphi,A=0)=0$), and for $E=0$,
\begin{equation} \label{eq:currwire}
  I(x,t)=-G\dot{\varphi}(x,t)=\frac{G}{e}[\mu_L-\mu_R](x,t)
  \quad,
\end{equation}
where $G$ now has the interpretation of a \textit{longitudinal
  conductance}. If all quasi-particles in the wire have integer electric
charge then $G$ is an integer multiple of $e^2/h$; (see
\cite{vwebeen}\cite{alchefr}). 

If there are tunnelling processes mixing left- and right-movers, due e.g. to
\textit{impurities} in the wire, then the term $\tilde{\Gamma}(\varphi,A)$ on
the R.S. of (\ref{eq:effac2}) does not vanish, even if $E=0$. The general
expression for the current $I$ in the wire is given by the equation
\begin{equation} \label{eq:currwire1}
  I(x,t)=\frac{G}{e}[\mu_L-\mu_R](x,t)+
         \frac{\delta\tilde{\Gamma}(\varphi,A)}{\delta A_1(x,t)}
  \quad,
\end{equation}
with $e\dot{\varphi}=\mu_L-\mu_R$ and $\dot{A}_1=E$. If scattering at the
impurities converts left- into right-movers, and conversely, the second term
on the R.S. of eq. (\ref{eq:currwire1}) does not vanish even if $E=0$, and
hence conductance is not quantized, anymore, in accordance with
experiment. However, charge transport over long periods of time still exhibits
``quantization'', provided $E=0$, due to the periodicity of $\tilde{\Gamma}$
in $\varphi$.

A more detailed account of our results and an analysis of the Luttinger
liquids in quantum wires in the presence of impurities will be given elsewhere.

\section{A 5-dimensional analogue of the Quantum Hall Effect, and Primeval
  Magnetic Fields in the Early Universe}
  \label{sec:5d}

Imagine, for a moment, that our world corresponds to a stack of $3$-branes
in a $5$-dimensional space-time. We suppose that all electrically charged
modes propagating through the $5$-dimensional bulk have a large mass
(comparable, e.g., to the Planck mass) and have parity-violating dynamics. We
may then ask whether there is an analogue of the quantum Hall effect in the
$(4+1)$-dimensional bulk. To be specific, we assume that there are two
parallel, flat $3$-branes separated by a $(4+1)$-dimensional slab $\Lambda$ of
width $\lambda$ representing the bulk of the system. Let $\hat{A}$ denote the
$5$-dimensional electromagnetic vector potential and $A$ the restriction to
the boundary, $\partial\Lambda$, of the slab of the components of $\hat{A}$
parallel to $\partial\Lambda$. Assuming that only the graviton and the photon
are massless modes, and dropping the gravitational contribution, the effective
action of such a system is given by 
\begin{eqnarray} 
  S^{\mathrm{eff}}_{\Lambda}(\hat{A}) & = &
    \frac{\varepsilon}{4\lambda}\int_{\Lambda}
    \hat{F}\wedge\ast\hat{F} \nonumber\\
  && +\frac{\sigma_T}{3} \int_{\Lambda}\hat{A}\wedge\hat{F}\wedge\hat{F} 
    \nonumber\\
  && +\Gamma_{\partial\Lambda}(A)+\;\mathrm{irrelevant}\;\mathrm{terms}
  \quad,\label{eq:5deffac}
\end{eqnarray}
with $\xi=(\xi^0,\xi^1,\ldots,\xi^4)\equiv(t,\underline{x},\xi^4)$, where $\xi^4$ is
the coordinate perpendicular to the boundary $3$-branes, which are located at
$\xi^4=0,\lambda$, respectively, and $\varepsilon$ is a dimensionless
constant. The first term on the R.S. of (\ref{eq:5deffac}) is a Maxwell term,
which is the dominant term, the second term is the $5$-dimensional
Chern-Simons action, the last term is the $4$-dimensional anomalous chiral
(boundary) action, which ensures that $S^{\mathrm{eff}}_{\Lambda}(\hat{A})$ is
gauge-invariant. From the theory of the chiral anomaly we infer that
\begin{equation} \label{eq:sigmatr}
  \sigma_T=\frac{e^3}{8\pi^2}\sum_{\alpha}q_{\alpha}^3
  \quad,
\end{equation}
where the $q_{\alpha}$'s are the charges of the chiral fermions propagating
along $\partial\Lambda$. The action $\Gamma_{\partial\Lambda}$ is the
$4$-dimensional version of the boundary action $\Gamma_{\partial\Lambda}$ in
eq. (\ref{eq:effac}). It is the generating functional of the Green functions
of chiral currents $j^{\mu}_{L/R}$ satisfying 
\begin{equation} \label{eq:5danom}
  \partial_{\mu}j^{\mu}_{L/R}\;=\;\pm\frac{\sigma_T}{3}
     \underline{E}\cdot\underline{B}
  \quad,
\end{equation}
where $(\underline{E},\underline{B})$ is the electromagnetic field on the boundary
$3$-branes. Modes of opposite chirality are localized on the two opposite
$3$-branes, (at $\xi^4=0$ and $\xi^4=\lambda$, respectively).

Imagine that the fields $\hat{F}_{\mu\nu}$, $\mu,\nu=0,1,2,3$ are
  \textit{independent} of $\xi^4$. We define the \textit{axion field},
  $\varphi$, by
\begin{equation} \label{eq:5axion}
  \varphi(\underline{x},t)=\int_0^{\lambda}A_4(t,\underline{x},\xi^4)d\xi^4
  \quad.
\end{equation}
After dimensional reduction to the boundary $3$-branes, $\partial\Lambda$, the
effective action in (\ref{eq:5deffac}) becomes
\begin{eqnarray} 
  S^{\mathrm{eff}}_{\Lambda}(\varphi,A) & \approx &
    \frac{\varepsilon}{4}\int_{\partial\Lambda}F_{\mu\nu}F^{\mu\nu}d^4x
    +\frac{\varepsilon}{2\lambda^2}\int_{\partial\Lambda}
         \partial_{\mu}\varphi\partial^{\mu}\varphi \;d^4x \\
    && +\sigma_T \int_{\partial\Lambda}\varphi F\wedge F \\
    && +\tilde{\Gamma}_{\partial\Lambda}(\varphi,A)\quad.\label{eq:5deffac1}
\end{eqnarray}
If tunnelling between the two boundary $3$-branes is suppressed completely the
boundary action $\tilde{\Gamma}_{\partial\Lambda}(\varphi,A)$ is independent
of $\varphi$ and can be combined with the 
Maxwell term to renormalize its coefficient. But if tunnelling processes
mixing fermions of opposite chirality are present then
$\tilde{\Gamma}_{\partial\Lambda}(\varphi,A)$ depends on $\varphi$ and is
$\neq0$ even if $F=0$. Tunnelling processes generate a small mass,
proportional to $Me^{-\lambda/l_P}$, of boundary fermions; (here $M$ is a
typical bulk mass scale, and $l_P$ is the Planck length). By repeating the
arguments explained in Section \ref{sec:constriction}, one finds that
$\tilde{\Gamma}_{\partial\Lambda}(\varphi,A)$ is \textit{periodic} in
$\varphi$ with period $\varphi_0$ proportional to $q_{\ast}^{-1}$, where
$q_{\ast}$ is the smallest electric charge of modes propagating along the
$3$-branes. 

We recall that $S^{\mathrm{eff}}(\varphi,A)$ is the generating functional of
the Green functions of the pseudo-scalar density and the electric current
density; in particular, $j^{\mu}=\delta S^{\mathrm{eff}}(\varphi,A)/\delta
A_{\mu}$. Plugging the expressions for
$j^{\mu}=\big{<}\mathcal{J}^{\mu}\big{>}$ and for
$\big{<}\bar{\psi}\gamma^5\psi\big{>}$ obtained from (\ref{eq:5deffac1}) into
Maxwell's equations and the equations of motion for the axion field, we find
the following equations of motion:
\begin{eqnarray} 
  F_{[\mu\nu;\sigma]} & = & 0 \quad, \nonumber \\
  F^{\mu\nu}_{;\sigma} & = & 2\beta\sigma_T(\varphi\tilde{F}^{\mu\nu})_{;\nu}
         +\beta\frac{\delta\tilde{\Gamma}(\varphi,A)}{\delta A_{\mu}} \quad, 
         \label{eq:fieldeq} \\
  \Box\varphi\equiv\varphi^{,\mu}_{;\mu}& = &
      -\beta'\lambda^2\left[
           \sigma_TF\tilde{F}+
           \frac{\delta\tilde{\Gamma}(\varphi,A)}{\delta \varphi}+
           k\mathrm{tr}(R\tilde{R})\right] \quad,\nonumber
\end{eqnarray}
where $\beta$ and $\beta'$ are dimensionsless constants, and the term
$k\mathrm{tr}(R\tilde{R})$, where $R$ is the Riemann tensor, comes from a term
$k\int\varphi\mathrm{tr}(R\tilde{R})$ in the effective action describing the
coupling of the axion to space-time curvature (which has not been displayed in
eq. (\ref{eq:5deffac1})). If there exist magnetic monopoles the first equation
in (\ref{eq:fieldeq}) must be replaced by
$F_{[\mu\nu;\sigma]}=j^M_{\mu\nu\sigma}$, where $j^{M}$ is the magnetic
  current 3-form. In conventional vector analysis notation,
  eqs. (\ref{eq:fieldeq}) take the form
\begin{eqnarray}
  \underline{\nabla}\cdot\underline{B} & = & 0 \nonumber \\
  \underline{\nabla}\wedge\underline{E} & = & \underline{\dot{B}} \nonumber \\
  \underline{\nabla}\cdot\underline{E} & = &
    2\beta\sigma_T\underline{\nabla}\varphi\cdot\underline{B} +
    \beta\frac{\delta\tilde{\Gamma}(\varphi,A)}{\delta A_0} \nonumber \\
  \underline{\nabla}\wedge\underline{B} & = &
    -\underline{\dot{E}}+\sigma_L\underline{E}
    +2\beta\sigma_T[\dot{\varphi}\underline{B}+
                    \underline{\nabla}\varphi\wedge\underline{E}]
    +\beta\frac{\delta\tilde{\Gamma}(\varphi,A)}{\delta\underline{A}}
      \nonumber \\
   \Box\varphi & = &
   -\beta'\lambda^2\left[2\sigma_T\underline{E}\cdot\underline{B}
     +\frac{\delta\tilde{\Gamma}(\varphi,A)}{\delta\varphi}
     +k\mathrm{tr}(R\tilde{R})\right] \quad, 
       \label{eq:fieldeq1}
\end{eqnarray}
where, in the fourth equation of (\ref{eq:fieldeq1}), the term
$\sigma_L\underline{E}$ has been added to describe a dissipative current
parallel to $\underline{E}$, with $\sigma_L$ the longitudinal conductivity;
(Ohm's law).

It is clear from eq. (\ref{eq:5axion}) that the time derivative,
$e\dot{\varphi}$, of the axion field has the interpretation of a (space-time
dependent) \textit{difference of chemical potentials} of right-handed and
left-handed charged modes propagating on the ``upper'' and the ``lower''
brane, respectively.

Absorbing the leading $A$-dependent contribution to
$\tilde{\Gamma}(\varphi,A)$ into a renormalization of the constant $\beta$,
the leading term in $\tilde{\Gamma}(\varphi,A)$ is independent of $A$ and has
the form 
\begin{equation} \label{eq:effaclead}
  \tilde{\Gamma}(\varphi,A)=\int U\left(\varphi(x)\right)d^4x \quad,
\end{equation}
where $U$ is a (temperature-dependent) \textit{periodic} function with period
$\varphi_0$. Plugging eq. (\ref{eq:effaclead}) into (\ref{eq:fieldeq1}), we
find that a special solution of (\ref{eq:fieldeq1}) is given by
$\underline{E}=\underline{B}=0$ and $\varphi$ solving the the equation
\begin{equation} \label{eq:axifieldeq}
  \Box\varphi=-\beta'\lambda^2\left[U'(\varphi)+k\mathrm{tr}(R\tilde{R})\right]
  \quad.
\end{equation}
As mentioned above, $U$ actually depends on the the temperature of the
universe: $U\approx0$ at temperatures well above the electro-weak phase
transition; while, at temperatures below the electro-weak scale, $U$ is a
non-constant, periodic function of $\varphi$ with minima at
$\varphi=n\varphi_0$, $n\in\mathbf{Z}$. Thus, at the time $t_{\ast}$ of the
electro-weak phase transition, the configuration
$\varphi(t_{\ast},\underline{x})$ corresponds, approximatively, to a solution
of 
\begin{equation} \label{eq:axifieldeq1}
  \Box\varphi=-\beta'\lambda^2\mathrm{tr}(R\tilde{R})
  \quad,
\end{equation}
and there is no reason why $\varphi(t_{\ast},\underline{x})$ should be close
to a minimum of the function $U(\varphi)$, or why
$\dot{\varphi}(t_{\ast},\underline{x})$ should be small. The source term
$-\beta'\lambda^2\mathrm{tr}(R\tilde{R})$ on the R.S. of
(\ref{eq:axifieldeq1}) does not vanish, provided there are gravitational waves
propagating through the universe. For a Friedman universe, it is proportional
to the amplitude of gravitational waves: thus, such waves can, in principle,
feed the growth of the axion field.

At times $t>t_{\ast}$, the equation of motion of the axion is given by
\begin{equation} \label{eq:axifieldeq2}
  \Box\varphi=-\beta'\lambda^2\left[U'(\varphi)+k\mathrm{tr}(R\tilde{R})\right]
  \quad,
\end{equation}
with $U\neq$constant. Assuming that gravitational waves eventually disperse
away, the term proportional to $\mathrm{tr}(R\tilde{R})$ will approach $0$, for
a Friedman universe. Let us suppose that, after inflation,
$\varphi(t,\underline{x})\approx\varphi(t)$ varies slowly over space. Then
eq. (\ref{eq:axifieldeq2}) reduces to an ordinary differential equation
\begin{equation} \label{eq:axifieldeq3}
  \ddot{\varphi}=-\beta'\lambda^2U'\left(\varphi(t)\right)
  \quad,
\end{equation}
to be solved for essentially random initial conditions,
$(\varphi(t_{\ast}),\dot{\varphi}(t_{\ast}))\neq(n\varphi_0,0)$,
$n\in\mathbf{Z}$. Eq. (\ref{eq:axifieldeq3}) is the equation of motion for a
pendulum in a potential force field, $-U'\left(\varphi(t)\right)$. Solutions
of (\ref{eq:axifieldeq3}) are given by
\begin{equation} \label{eq:axifieldeq4}
  \varphi(t)=\frac{\mu_L-\mu_R}{e}t+\alpha(t)
  \quad,
\end{equation}
where $\alpha$ is a periodic function of $t$.

Next, we linearize eqs (\ref{eq:fieldeq1}) around the special solution
$\underline{E}=\underline{B}=0$, $\varphi(t)$ as in
(\ref{eq:axifieldeq4}). This is not a difficult task. One finds that for
sufficiently small wave vectors, $\underline{k}$,
($|\underline{k}|<2\beta\sigma_T(\mu_L-\mu_R)/e$, for $\alpha=0$), there are
\textit{exponentially growing transverse modes},
$\underline{\hat{B}}(\underline{k},t)$, of the magnetic field with
non-vanishing magnetic helicity. One expects that axion field configurations
which are slowly varying in space lead to qualitatively similar
instabilities. When combined with the galactic dynamo mechanism they might
provide an explanation of the large-scale magnetic fields observed in the
universe; (but see \cite{giovagraru} for discussion of some of the
difficulties with this and other scenarios). We hope to present a more
detailed account of our results, in particular of the possible r\^{o}le of
gravitational waves, elsewhere.

\textbf{Acknowledgments}.
We thank G. M. Graf for explaining to us the notion of a ``quantum pump'',
\cite{gm}, and I. I. Tkachev and Ph. Werner for very valuable discussions on
the material presented in Section \ref{sec:5d}.

\end{document}